\documentstyle[aas2pp4,psfig]{article}

\slugcomment{To appear in AJ number NN}
 
\lefthead{Rosa A. Gonz\'alez et al.}
\righthead{Galactic Extinction}
 
\def\arcsper{\ifmmode \rlap.{^{\prime\prime }}\else
$\rlap{.}{^{\prime\prime}} $\fi}
\def\arcmper{\ifmmode \rlap.{^{\prime }}\else $\rlap{.}{^\prime} $\fi}
\def\degper{\ifmmode \rlap.{^{\circ }}\else $\rlap{.}{^\circ} $\fi}
\def\asec{$^{\prime \prime}$}

\def\deg{\ifmmode^\circ\else$^\circ$\fi}

\def\H2{H$_2$}

\def\13co{$^{13}$CO}
\def\nh3{NH$_3$}

\def\gax{${_>\atop^{\sim}}$}

\def\cm2{cm$^{-2}$}
\def\cm3{cm$^{-3}$}

\def\samename{\vrule height0.4pt depth0.0pt width1.0in \thinspace.}
\def\sref#1 #2 #3 #4 {#1, #2, #3, #4}

\begin{document}

\title {GALACTIC EXTINCTION
FROM COLORS AND COUNTS \\OF FIELD GALAXIES IN WFPC2 FRAMES:\\[0.2in]
An Application to GRB~970228\altaffilmark{1}}
 
\author {Rosa A.\ Gonz\'alez and Andrew S. Fruchter,}
\affil {Space Telescope Science Institute, Baltimore, MD 21218; 
ragl \& fruchter @stsci.edu}

\footnotesize\vskip 0.25in{
\centerline {AND}}

\author {Boris Dirsch}
\affil {Sternwarte der Universit\"at Bonn, D-53121 Bonn, FRD;  
bdirsch@astro.uni-bonn.de}
\altaffiltext{1}
{Based on observations with the NASA/ESA Hubble
Space Telescope obtained at the Space Telescope Science Institute, which is
operated by the Association of Universities for Research in Astronomy, 
Incorporated, under NASA contract NAS5-26555.}
 
\begin{abstract}
We develop the ``simulated extinction method''
to measure average foreground Galactic extinction
from field galaxy number-counts and colors.
The method comprises simulating extinction in suitable reference
fields by changing the isophotal detection limit.
This procedure  
takes into account selection effects,
in particular, the change in isophotal detection
limit (and hence in isophotal magnitude completeness limit)
with extinction, and the galaxy color--magnitude relation.

We present a first application of the method to
the HST WFPC2
images of the gamma-ray burster GRB~970228.
Four different WFPC2 high-latitude fields, including
the HDF, are used as reference 
to measure the average extinction towards the GRB in the $F606W$ passband.
From the counts, we derive an average extinction of
$A_V$ = 0.5 mag, but the dispersion of 0.4 mag 
between the estimates from the different reference
fields is significantly larger than can
be accounted by 
Poisson plus clustering 
uncertainties. 
Although the counts differ, the average colors of the field galaxies 
agree well. The extinction implied by the average color difference
between the GRB field and the reference galaxies is $A_V$ = 0.6 mag, with a
dispersion in the estimated extinction from the 
four reference fields of only 0.1 mag.
All our estimates
are in good agreement
with the value of 0.81$\pm$0.27 mag obtained by
Burstein \& Heiles, and 
with the extinction of 0.78$\pm$0.12 measured by Schlegel et al.\ from
maps of dust IR emission. 
However, the discrepancy between the widely varying counts and the
very stable colors in these
high-latitude fields is worth investigating.

\end{abstract}


\newpage

\normalsize

\section{INTRODUCTION}\label{intro}

\subsection{Historical background} \label{history}

Since the classical work of Trumpler (1930a,b) produced
strong evidence of the existence of interstellar
dust, several methods have been devised to try to measure the
Galactic extinction towards extragalactic objects in
a reliable way. 
In particular, there is a long history on the use
of background galaxies to probe extinction.
As early as 1934, Hubble 
realized that the
variation of galaxy number-counts with Galactic latitude
was well represented by the cosecant law, and 
was therefore consistent with a layer of
absorbing material of equal thickness above and below the
Galactic plane. 
In 1958, Holmberg found that the 
observed colors (corrected for redshift and internal extinction) of 
spirals of different types
correlated with Galactic latitude.
In 1967, Shane and Wirtanen 
published the best sample up to that date of galaxy number-counts;
their results showed conclusively that galaxies were clustered,
but also that Galactic extinction was patchy.

When compared to the colors of globular clusters and early-type 
field galaxies,
the Shane-Wirtanen number-counts also raised the possibility
of neutral or grey extinction 
(i.e., $R_V = A_V / E(B - V) >>$ 3),
especially close to the Galactic plane. 
Given that this ``grey'' extinction could be an artifact
of comparing  
total extinction derived from what we cannot see 
with selective extinction
(reddening) inferred from what we can see, 
Peterson (1970) used the observed colors and magnitudes of
brightest cluster ellipticals 
in an attempt to measure both extinction and reddening from the same objects. 
Sandage (1973,1975) employed the same technique on 
brightest cluster members and radio galaxies projected on the Galactic plane; 
he measured $R_V \sim$ 3, but of course
the simple fact
that he could observe the cluster galaxies indicated that these were 
regions of low extinction in the plane of the Galaxy. 

In a series of papers, Heiles (1976), and Burstein \& Heiles
(1978a,1978b) discussed the relationships between HI column density,
the Shane-Wirtanen galaxy counts, 
and reddening. Heiles (1976) also addressed
the influence of surface brightness profile and distance
(given the dependence of apparent angular diameter
on distance) on the probability of galaxy
detection. Burstein \& Heiles (1978b) elaborated on the
effects of variable background when comparing number-counts; 
they also 
concluded that the (Poisson plus clustering) variance 
in the Shane-Wirtanen number-counts was enough
to prevent an accurate determinantion of extinction
towards the Galactic poles.
The end result of their work was Burstein \& Heiles (1982), 
the primary reference for estimating Galactic extinction
towards extragalactic objects until the recent work of
Schlegel, Finkbeiner \& Davis (1998) using the IRAS and COBE/DIRBE
(Diffuse Infra Red Background Experiment) 100 $\mu$m maps.

Also very recently, Gonz\'alez et al.\ (1998, hereafter G98) 
have used galaxy counts and
colors to measure the total opacity of spiral galaxy disks. 
This is technically a much more complicated problem, 
owing to the crowding and confusion from stars, star clusters, and 
HII regions in the foreground disk. 
In order to decouple the effects of confusion 
from those of extinction, we developed the ``synthetic field'' method:
suitable and properly scaled reference fields
(the Hubble Deep Field, HDF, in the case of G98) are added directly
into the spiral galaxy images; 
the reference frames are attenuated to mimic different amounts
of extinction and reddening, until one recovers the same number of 
simulated galaxies as there are real
background galaxies.
Relative to G98, this paper is a simpler application 
of extinction simulations using reference fields.
However, the present work also represents an advance 
in the treatment of the color information, 
and also because we use additional reference fields.

In general, the ability to determine total average
extinction from galaxy number-counts is 
limited by clustering, and 
by the difficulties in assessing completeness. Galaxy colors,
on the other hand, are biased towards regions of low opacity (since 
observing the galaxies is a pre-condition to measure their
colors).
However, if it can be safely assumed that
the region of interest
can be sampled everywhere with field galaxies,
it is possible to determine an extinction
law from the combination of colors and counts
(G98) or, assuming an extinction law
{\it a priori}, it is feasible to derive total extinction
from the reddening of the galaxies alone. 
This is what Zaritsky (1994) attempted to do in the
halos of external galaxies, through the comparison of the
colors of all background galaxies detected with the colors
of galaxies in control fields.  But when field galaxies are  
thrown in together regardless of type, a new problem
emerges: galaxies have a color--magnitude dependence
(Tyson 1988, Williams et al.\ 1996). This means that
the color offset has to be measured with respect to
samples that have the same magnitude limit as the
{\it extinction-corrected} limit of the galaxies detected
in the field of interest.  Zaritsky (1994) 
used a maximum-likelihood method to circumvent 
this problem. We have developed a much
simpler and more intuitive method to account for
the color--magnitude relation when 
deriving Galactic extinction from 
the colors of field galaxies. 

\subsection{The simulated extinction method.} \label{method}

In this paper, we describe the ``simulated extinction method''   
to measure foreground Galactic extinction 
from background field galaxy number-counts and colors.
The method takes into account selection effects,
in particular, the change in isophotal detection
limit (and hence in isophotal magnitude completeness limit) 
with extinction, and the galaxy color--magnitude relation.
We simulate 
extinction in suitable reference
fields by changing the isophotal detection limit. 
The procedure is repeated for different values of extinction.
To make use of the cumulative number-counts, a plot is drawn of the
total number of galaxies recovered in each simulation vs.\ the assumed extinction in
order to ascertain the ``best'' fit to the actual number-counts 
in the field of interest (G98).
To use the colors, we measure the difference between the average color of 
the galaxies in the field of interest, and the average color of
the reference galaxies as we change the isophotal
detection limit of the latter. Note that we are {\em not} simulating the
reddening of the reference fields, just the change of average color
with isophotal detection limit (the galaxy color--magnitude relation).
Next, we plot the color difference vs.\ ``extinction'',
or detection threshold. The
extinction in the field of interest is given by the intersection
of this function (which decreases with extinction or a brighter
isophotal detection limit) with the Galactic reddening law.

We present a first application of the method to 
the Hubble Space Telescope (HST) Wide-Field Planetary Camera 2 (WFPC2) 
images of the gamma-ray burster GRB~970228.
The extinction towards the GRB has recently been a matter
of contention (Castander \& Lamb 1998); 
in the absence of a measured redshift of the host,
the assumed unreddened colors determine the
plausibility of the object being in the Milky Way or extragalactic. 
As we will show, both galaxy counts and 
(assuming a Galactic reddening law; Schlegel et al.\ 1998) 
galaxy colors
yield independently the same, relatively low ($A_V \sim$ 0.6 mag)
foreground extinction towards the burster.

\section{THE REFERENCE FIELDS}

As discussed elsewhere (G98),
HST WFPC2 
images offer a significant advantage over images from
ground-based telescopes in their greatly improved resolution, which 
facilitates the separation of stellar objects from
galaxies. Ideally, reference fields should be available
with the same camera and range of colors, and at least
as deep or deeper in exposure time as the images of interest.
Obviously, the extinction towards the reference fields should be 
known (or assumed). 
Fortunately, there are by now
many fields observed with the WFPC2 which can satisfy these criteria,
including the Medium Deep Survey fields (typically exposures of
\gax 5000 s in the $F606W$ and $F814W$ passbands;
Griffiths et al.\ 1994; Ratnatunga, Griffiths, \& Ostrander 1998), the
field around the weak radio-galaxy 53W002
(24 orbits in 3 colors; e.g., Pascarelle et
al.\ 1996a,b), the two deep Westphal fields
($\sim$ 25000 s each in both $F606W$ and $F814W$; Westphal, Kristian, 
\& Groth 1994) and, of course, the
HDF (150 orbits in 4 colors, Williams et al.\ 1996),
which offers the deepest set of images. 

The counts and colors of the
galaxies in the reference images should be ``representative'' 
of galaxies in the field or at least
well characterized. However, this is not as crucial as it
might seem; as discussed in the conclusions to G98,    
the relevant quantities for this kind of work are the dispersions
of the galaxy number-counts and colors, and not the means.
Even if we manage to 
determine very accurately the means of 
these properties 
in a typical WFPC2 reference field,
the larger source of error will still be the
``single realization'' of the field
that we are investigating.
Under these circumstances, the dispersion is a 
fair estimator of the expected error
in the derived extinction.

For this work, we use as reference fields the HDF, the field around
the weak radio-galaxy 53W002, and the two deep Westphal fields.
Like the data on the GRB~970228, each reference image 
comprises 3 contiguous\ \ \ Wide\ \ \ Field\ \ \ Camera\ \ \ fields,\ \ \ 
plus\ \ \ one\ \ \  
Planetary\ \ \ Camera\ \ \ (PC)\ \ \ field\ \ \ which\ \ \ we\ \ \ do\ \ \ 
not\ \ \ use\ \ \ here 
\footnote{Actually, the optical counterpart of the gamma-ray
burster was imaged on the PC field.}.
Table 1 lists the equatorial coordinates, 
Galactic coordinates, and exposure times in 
the $F606W$ and $F814W$ passbands of  
the reference fields, and of the GRB~970228 images as well.
The Burstein \& Heiles (1982) $E(B - V)$ 
towards all of the reference fields is basically zero, but we assume
$E(B - V) = 0.02$, from the systematic offset found
by Schlegel et al.\ (1998). 
We choose to estimate the clustering
error from the literature (Roche et al.\ 1993; Brainerd et al.\ 1995), 
because we would need a much bigger
sample of reference WFPC2 fields to measure the error directly
from the data. 

\section{OBSERVATIONS AND DATA REDUCTION} \label{datared}

We use the Version 2 Release 
images of the HDF (Williams et al.\
1996) retrieved from the STScI/HST archive. This release of the HDF was
made from dithered\ \ exposures\ \ using\ \ the\ \ ``drizzling''\ \
method\ \ (Fruchter \&
Hook 1997).  The HDF was observed in  
the 4 photometric bands designated $F300W$, $F450W$, $F606W$, and $F814W$ 
(cf.\ the HST Data Handbook); we only use the images in the
$F606W$ and $F814W$ passbands for this investigation (Table 1).

Like the public release HDF images, 
the WFPC2 images of the fields surrounding GRB~970228 and 53W002, as
well as the two Westphal fields, were
drizzled onto  sub-sampled output images; care was taken
to align the $F606W$ and $F814W$ images of each field.  A final
output pixel scale of $0\farcs05$ was used, rather than the $0\farcs04$
of the HDF. (As the HDF was the first application of drizzling, a very
conservative pixel scale was used.)  Additionally, the images were
cleaned of blemishes using the ``blot'' task (Fruchter \& Hook 1997).  
Since ``blot''
is only used to produce comparison images for the creation of blemish
masks, its use should have no effect on the photometry of the final
images.  Similarly, we are able to adjust for the effect of the differing
pixel scales of the HDF and the other images by altering the source extraction
software parameters (\S\ref{extraction}).

\section{ANALYSIS AND RESULTS} \label{datanalysis}

When using an isophotal brightness threshold to   
detect and perform photometry of galaxies,
extinction has a double effect. 
It will reduce the surface brightness and the 
total apparent luminosity of the galaxies, of course,
but it will also shift a fixed isophotal detection
threshold to intrinsically brighter magnitudes. 
On average, the
difference between the measured isophotal magnitudes
and the total apparent magnitudes of galaxies 
will be larger than in the absence of extinction, 
because the isophotal brightness will be 
integrated over a smaller area of the galaxy
(Shane \& Wirtanen 1967; Heiles 1976).
In other words, 
the differential number-count vs.\  
apparent magnitude relation 
will shift towards fainter magnitudes by an amount 
larger than the extinction. 
Another problem with comparing differential 
counts vs.\ magnitude
in two fields is that they 
will be influenced
(mainly at brighter magnitudes, where galaxy clustering is worst)
by overdensities associated with particular large scale structures.
A much better approach is to use cumulative counts, taking into
account selection effects.

\subsection{Object extraction} \label{extraction}

We used SExtractor (Bertin \& Arnouts 1996) to perform the 
photometry on all objects in the GRB field,\ \ 
up\ \ to\ \ the\ \ $25\ \ {\rm
mag}_{AB}$\ \ arcsec$^{-2}$\ \ isophote\ \ in\ $V_{F606W}$.
This threshold is somewhat conservative, but it
allowed the detection of most objects that we could
identify as galaxies by eye, and at the same time
avoided the identification of noise as spurious galaxies.
The source detection\footnote{A gaussian smoothing kernel with a
FWHM of 2\asec was applied in all cases.}, object deblending, and object
classification (stars vs.\ galaxies)
were done in the $F606W$ image because
it is deeper than the $F814W$ frame (Table 1). Also, fainter field galaxies
are statistically bluer, so searching in $F606W$ 
allows the detection of more objects.
Once galaxies had been identified in the $F606W$ data, 
their ($F606W$ - $F814W$) color was measured  
in a fixed aperture of 0\arcsper4 diameter (\S\ref{colors}). 
It is worth emphasizing that separation between stars and
galaxies is not a serious problem at the resolution and 
depth of the HST images. Not only the separation on the
basis of morphology is easier; in addition, 
owing to the finite size of the Milky Way,
at fainter magnitudes the number-counts 
of Galactic stars are drastically reduced.
We have estimated that the error in number-counts 
introduced by uncertainties in object separation is, at most, $\sim 2\%$,
about half of the Poisson error. 

\subsection{Extinction from galaxy counts} \label{counts}

Following closely on the ideas developed in
G98, photometry of galaxies was performed in the
control fields at different simulated extinctions,
i.e., at progressively brighter detection thresholds,
starting with the 25 $V_{F606W}\ {\rm
mag}_{AB}$ arcsec$^{-2}$ isophote
adequate for the GRB image, and at intervals
of 0.40 mag in $F606W$ (or 0.45 mag in $V$; Schlegel et al.\ 1998).
This will take into account the dependence of
the cumulative number-count vs.\ extinction curve on object selection
biases.
With our procedure, the higher signal--to--noise ratio of the reference
fields (especially the Hubble Deep Field) will result in a
smaller photometric error, but will have no impact on the
number of galaxies detected.
We tested this assumption by adding noise to the HDF, 
until it had the same signal-to-noise
ratio as the GRB image; the search on this degraded image 
yielded 277 instead of 275 galaxies (Table 2).

\begin {figure}
\hskip 0.7in\psfig{figure=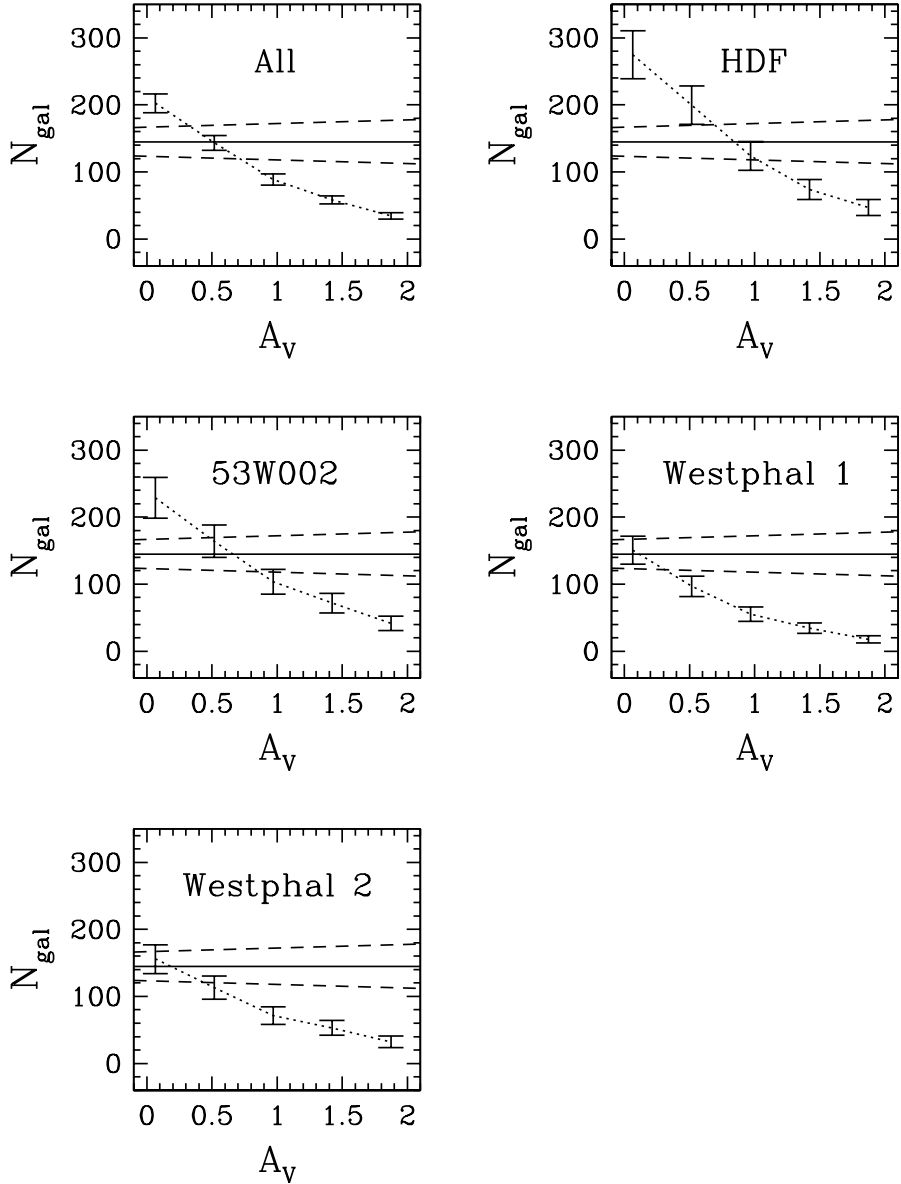,width=5in,angle=0}
\caption{{\em Average number of recovered galaxies
vs.\ simulated extinction of $A_V$. Top left:}
All control fields; {\em top right:} HDF;
{\em middle left:} 53W002; {\em middle right:} Westphal 1;
{\em bottom left:} Westphal 2. 
The error bars include the statistical Poisson uncertainties,
as well as the field-to-field variations expected from
galaxy clustering.
{\em Horizontal solid line:} number of galaxies
in the GRB~970228 field,
with its uncertainty ({\em dashed lines}).
For fixed number-counts, the error 
from galaxy clustering would increase
at higher extinctions because the completeness limit
would move towards intrinsically brighter magnitudes,
where clustering is worse.}
\label{fcounts}
\end{figure}

Figure\ \ref{fcounts} shows the comparison between the number of 
galaxies found in the GRB~970228 field (horizontal line) 
and the number of
``control'' galaxies recovered in the
reference fields at each different simulated extinction.
The actual measurement
of the extinction toward the gamma-ray burster field was 
performed by linearly interpolating
between the simulations with numbers of recovered galaxies
immediately above
and below the number of galaxies in the GRB field. 
Panel 1 compares the galaxies in the
field of the burster to the average of all 4 control fields; 
panels 2 through 5 show the results for the different 
reference fields.

The error bars include the Poisson counting error 
and the galaxy clustering error
expected from the two-point correlation function
$w(\theta = 139^{\prime\prime})$\footnote{
The area of the three Wide Field Camera chips is equal to
a square of 139$^{\prime\prime}$ on the side. Since the
field of view is not square, this slightly overestimates
the value of the correlation function.} 
(Roche et al.\ 1993; Brainerd et al.\ 1995).
The correlation function depends on the 
completeness limit of the galaxies; in the
case of the above mentioned references, $w(\theta)$ is 
measured in the $R$-band. 
We have used the HDF images to determine 
the number-count completeness limit
to our surface brighteness detection threshold of 25 
mag$_{AB}$ arcsec$^{-2}$ in $F606W$.
From histograms 
of differential counts vs.\ aperture magnitude in increasingly
larger apertures, we find that,
in the absence of extinction, the number-counts are complete to
a total integrated AB magnitude of 25.5 in
$F606W$.
This is equivalent to an $R$ (apparent) brightness
of $\sim$ 25 mag\footnote{
We have used the IRAF/STSDAS (Tody 1993) program {\em synphot} to determine
that for a K2 giant, which has ($F606W - F814W$)$_{AB}$ = 0.5, that is, the
same color as the average of all the reference galaxies to
our detection limit, ($F606W_{AB} - R_{Cousins}$)
= 0.4 mag.
Reassuringly, according to Smail et al.\ (1995), 
galaxies with a total $R$ apparent magnitude between 23.0 and 25.5 
have ($V - R$) $\sim$ 0.6 mag, or ($F606W_{AB} - R_{Cousins}$) = 0.4.
We roundoff these estimates 
to 0.5 mag, given that the references used for the two-point
correlation function provide their measurements in half-magnitude
steps. If anything, this slightly overestimates the 
expected clustering error. 
}.  As we change the detection threshold
to simulate extinction in the control fields, we move the 
completeness limit accordingly, towards brighter magnitudes.

Table 2 shows, for each reference field and for different
simulated extinctions in $V$, the assumed  
$R$ magnitude completeness limit, 
$w(\theta = 139^{\prime\prime})$, and the number of detected
galaxies with their total 1$\sigma$ error; 
in parenthesis, we list separately the Poisson counting error and
the galaxy clustering uncertainty. 
Due to rotation between the $F606W$ and the $F814W$ observations  
of the GRB~970228,
all the reference fields have an area a few percent larger than
that of the GRB mosaic; the number-counts in Figure \ref{fcounts} and
Table 2 have been adjusted for these differences.
The bottom part of the table displays the 
number-counts measured in the GRB~970228 field, 
as well as the extinctions in $V$ derived from their comparison 
against the counts in the control fields.
The quoted extinctions include the additional 0.07 mag in $A_V$
implied by the systematic offset in $E(B - V)$, relative to
Burstein \& Heiles (1982), found
by Schlegel et al.\ (1998). 
Here also, we give the total error in the derived extinction and,
in parenthesis, the error from shot-noise in the number-counts and
the contribution to the error from galaxy clustering. 

From the average of the number-counts in all the control fields, 
we measure an extinction of
$A_V = 0.51\pm0.22$ mag.
We notice, however, that the dispersion of the 
number-counts in the few WFPC2 reference fields we use is 
about twice the estimated error bars; for reasons that are
yet to be understood, the two Westphal fields have 
fewer galaxies than the HDF and the 53W002 field, by an
amount that is significantly larger than the expected clustering error. 
Comparison of the counts in the GRB field with those of the HDF and
the 53W002 field gives a best estimate of the extinction
of $A_V$ = 0.75$\pm$0.24; 
if the two Westphal fields are averaged in with the richer fields,
and we use the dispersion to estimate the error, 
the best value of the extinction is $A_V = 0.51\pm0.37$ mag.

\subsection{Extinction from galaxy colors}\label{colors}

We measured the average colors of the reference galaxies, 
once again at progressively brighter detection thresholds,
starting with the $25\ V_{F606W}\ {\rm
mag}_{AB}$ arcsec$^{-2}$ isophote
adequate for the GRB image. 
Following G98, 
after identification in the $F606W$ images,
we measured the color of each individual
galaxy in a fixed aperture, in this case of
0\arcsper4 diameter. The goal is to avoid contaminating
the measured galaxy colors with the color of the sky.
This means that the main effect of the change in detection threshold
(and therefore in isophotal magnitude limit) 
is the loss of faint (statistically bluer) galaxies. 
The error introduced by not measuring
possible color gradients over larger galaxy radii should be
negligible, especially because the measurement of a
color gradient depends very critically on an accurate estimation of the
background
\footnote{When using isophotal apertures, the average color 
of the galaxies in each field  
changes by less than the photometric error,
relative to the average color measured in the
0\arcsper4 apertures.}.
We assume that the error in the
mean color is its measured dispersion divided by $\sqrt{N}$, where
$N$ is the number of galaxies used for the color measurement
\footnote{Given that the color distribution of galaxies is not gaussian,
in G98 we compared the gaussian estimate of the error to the dispersion
in the mean color of 100 random
samples of $N$ 
HDF galaxies (where $N$ is the number of
galaxies revcovered at each simulated extinction); we found that the
error predicted from bootstrapping 
statistics is of the same order as the error found by assuming a
gaussian distribution.}. 

Since the errors in the average colors are of the order
of 0.05 mag (Table 3), in the simulations 
we incremented the detection threshold in
steps of 0.10 mag in $F606W$, or 0.11 mag in $V$ (Schlegel et al.\ 1998).
We remind the reader that we did {\em not} simulate the
reddening of the reference fields, just the change of average color
with isophotal detection limit (the galaxy color--magnitude relation).
We then calculated the color difference ($\Delta\ Color$) 
in $(F606W - F814W)$ between the reference
galaxies and the galaxies in the GRB field.

\begin {figure}
\hspace*{-0.1in} \psfig{figure=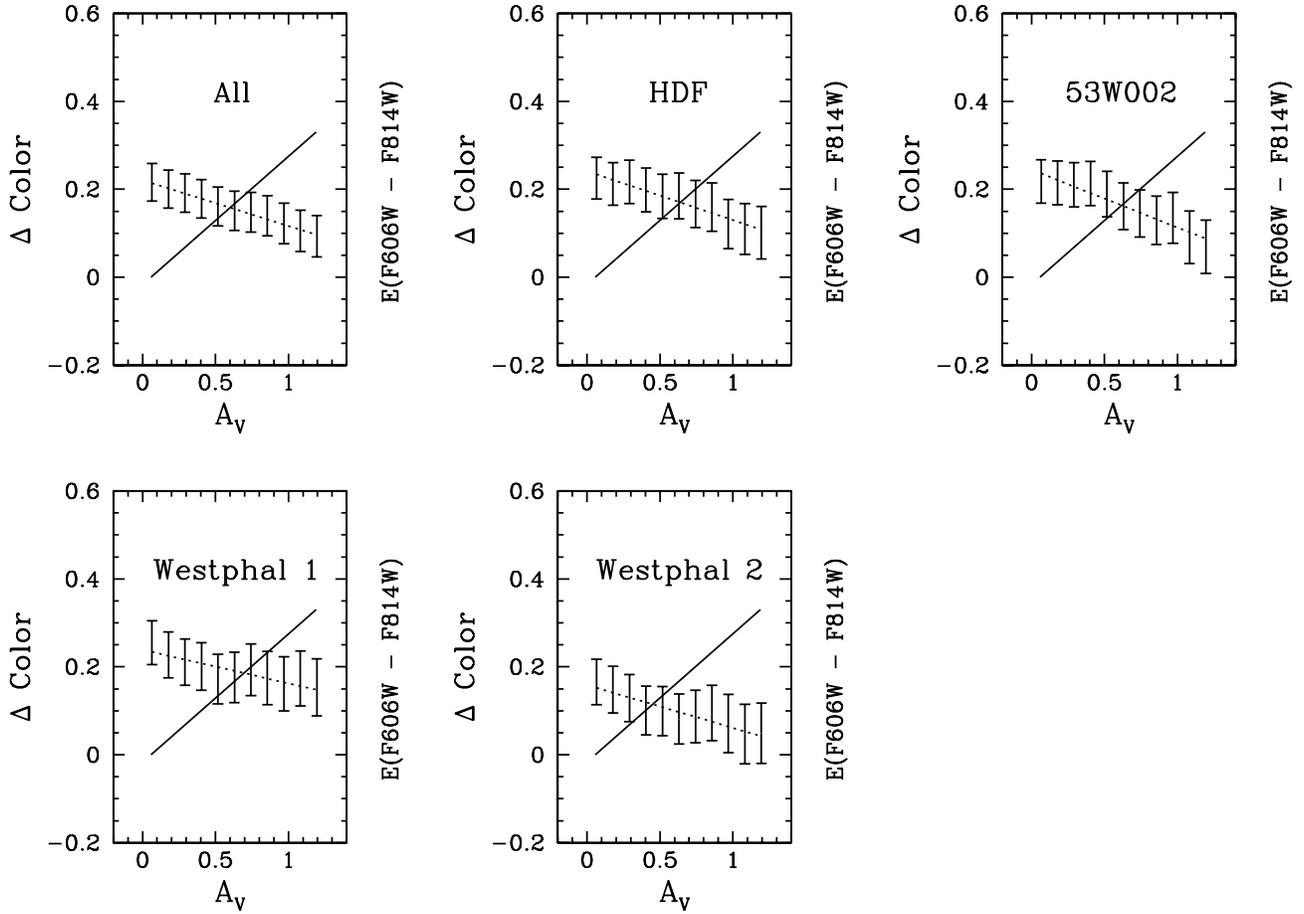,width=7in,angle=-90}
\caption{{\em Difference in average color between reference
galaxies and galaxies in the GRB field.
Top left:}
All control fields; {\em top middle:} HDF;
{\em top right:} 53W002; {\em bottom left:} Westphal 1;
{\em bottom middle:} Westphal 2.
The difference diminishes as the detection threshold 
moves to brighter magnitudes due to extinction;
the error bars assume a gaussian distribution of 
the colors.
{\em Dotted line:} Linear fit to the color difference;
{\em solid line:} Galactic reddening law (Schlegel et al.\ 1998).
The intersection of these two lines yields the extinction towards the
burster (see text).}
\label{fcolors}
\end{figure}

Figure \ref{fcolors} shows the change in 
$\Delta\ Color$ with isophotal detection limit. 
Panel 1 compares the galaxies in the
field of the burster to the average of all 4 control fields; 
panels 2 through 5 show the results for each one of the different
reference fields.
$\Delta\ Color$ decreases at brighter detection thresholds
(i.e., higher extinction), due to the color--magnitude
relation shown by field galaxies (brighter galaxies are
statistically redder). The errors are higher at high extinction,
both because fewer galaxies are detected and because the
intrinsic dispersion in color is higher for apparently brighter
galaxies. Figure \ref{fcolors} also displays the 
Galactic reddening line, that is, $E(F606W - F814W)$
vs.\ $A_V$ 
(Schlegel et al.\ 1998). The intersection between the two
relations measures the foreground Galactic extinction $A_V$
towards GRB~970228.
The observed average color of the galaxies in the GRB field
equals the sum of the control field average color plus the reddening
at the inferred extinction (cf.\ Table 3).
Albeit much simpler, our procedure is equivalent to 
{\em simultaneously} changing the
isophotal detection threshold {\em and} reddening the control galaxies
following a Galactic extinction law, until  
their average color is the same as that of the galaxies in the
GRB field. This is so because we base the galaxy detection 
in only one band, and we do not
use an upper limit in color as a galaxy selection criterion. 

Table 3 shows, for each reference field and for
different simulated extinctions in $V$, the average color of the
galaxies and the difference in color with respect to the galaxies in
the GRB field. The bottom of the table displays the measured color
of the galaxies in the GRB field, and the derived extinctions in $V$ from
the comparison with the galaxy colors in the reference fields. 
Accounting as described for the color--magnitude relation of field galaxies,
and assuming a foreground screen Galactic reddening law, 
a comparison between the 
average colors of, respectively,
the control galaxies and the galaxies in the GRB field 
yields $A_V$ = 0.61$\pm$0.11 mag; the error represents the
scatter between the measurements obtained from the different control fields.
In contrast to the number-counts, the galaxy colors 
in all 4 reference fields agree well; 
within the errors, the colors of all the fields are the
same, at all detection thresholds.

\section{SUMMARY}\label{summary}

We\ \ have\ \ developed\ \ the\ \ ``simulated\ \ extinction\  
method'' to
measure foreground Galactic extinction from the 
number-counts and colors of field galaxies.
The method simulates extinction in suitable reference
fields by changing accordingly the isophotal detection limit.
This procedure allows the calibration of selection effects and
systematics, in particular the galaxy color--magnitude
relation, and the change in isophotal
detection limit with extinction (and hence the change in both the
isophotal magnitudes of galaxies and in the number-count completeness limit).

We have used 4 HST WFPC2 reference fields 
to measure forward Galactic extinction 
towards the field of the gamma-ray burster GRB~970228.
Comparison of the counts in the GRB field with those of the HDF and
the 53W002 field gives a best estimate of the extinction
of $A_V$ = 0.75$\pm$0.24; if the two Westphal fields are
included, the value of the extinction is $A_V = 0.51\pm0.37$ mag.
Although the counts differ, the average colors of the field galaxies
agree well. 
The extinction implied by the average color difference
between the GRB field and the 4 reference fields is $A_V$ = 0.61 mag; the
dispersion in estimated extinction between the 
4 reference fields is only 0.11 mag.
However, there may be a slight bias 
in the value measured from the colors; if the extinction is
clumpy, the most obscured galaxies may fall out of the sample and
not affect the average color. Thus the estimate of $A_V$ = 0.61 mag may 
be a lower limit. 

\section{CONCLUSIONS AND DISCUSSION}\label{discussion}

All our measurements
are in good agreement
with the value of 0.81$\pm$0.27 mag obtained by
Burstein \& Heiles (1982; error re-estimated by Schlegel et al.\ 1998), and
with the extinction of 0.78$\pm$0.12 mag measured from
maps of dust IR emission (Schlegel et al.\ 1998).
Ultimately, all of these estimates are averages
with different spatial resolutions and over different
components of the ISM, at a low latitude position
where the Milky Way exhibits a steep gradient in extinction
(cf.\ Burstein \& Heiles 1982); none of these methods is capable
of measuring the extinction in the exact line of sight to the
GRB. 

It is very exciting that the colors of the
field galaxies agree so well, despite the big differences in
number-counts and
even though, for example, there are no clusters at low redshift
in the HDF, while the 53W002 field was selected around a weak radio
galaxy
(in no case did we try to identify and excise galaxy
clusters).
Moreover, comparison of the extinction-corrected color of
the GRB host to the ($F606W - H$) colors of the HDF galaxies
(Fruchter et al.\ 1998) suggests that the value of $A_V$ = 0.6 mag
derived from the colors might be closest to the correct extinction;
already with the value of $A_V$ = 0.75 mag adopted in that paper, 
the GRB host is as blue as the bluest objects in the HDF. 

But although the tightness in the colors is very
interesting {\it per se} and from the point of view of
their usefulness to probe extinction,
the discrepancy between the widely varying counts and the 
very stable colors in these
high-latitude fields is worth investigating.

The 53W002 field has a slightly lower exposure time than the two
Westphal fields. This, and the galaxy search performed in the
degraded version of the HDF (\S\ref{counts}) eliminate the possibility that
the discrepancy is caused by the different signal--to--noise ratios
of the reference fields.
The disparity occurs across all magnitude
bins fainter than 21.5 $V_{F606W}$ mag (isophotal), and 
applying a fainter detection threshold 
does not really help to reduce it.
We looked at the counts and colors of the
reference galaxies when using a surface
brightness detection limit of $26\ {\rm
mag}_{AB}$ arcsec$^{-2}$ isophote in $V_{F606W}$
(a threshold set by the 53W002 field, i.e., the shallowest 
of the reference images). The colors remain as stable;
the number-counts, on the other hand, start to converge,
but not nearly as quickly as the expected clustering
error drops.

It is possible that the problem 
lies in the quoted clustering errors.
To our knowledge, the field-to-field variance on 
the scale of the WFPC2 field of view has not
been investigated, even on ground-based data.
In fact, the
characteristics of fields of the size, spatial
resolution, and depth of our reference images are
a matter of intense study at present.
In this regard, although the properties of the HDF
(Villumsen, Freudling, \& Da Costa 1997)
appear to be consistent
with those derived for ground-based data
(Brainerd et al.\ 1995),
results on other deep fields are not yet available.
The clustering of galaxies brighter than $I$ = 23 mag in
prerefurbishment WFPC Medium Deep Survey images
(Neuschaefer et al.\ 1995)
is also consistent with
ground-based data (Roche et al.\ 1993), but
the completeness limit of the survey is much brighter
than that of the WFPC2 deep fields.

We could also have been too optimistic in our estimate
of the completeness limit.
If we disregard the results of the aperture photometry
and instead take the view of Tyson (1988),
that the completeness limit of a survey
is about two magnitudes brighter than the
surface brightness detection threshold, then our
number-counts are complete to only $R \sim$ 22.5 in the absence
of extinction. This will in fact bring our errors in line with
the observed dispersion of the data.

Lastly, 4 fields are not that many. We cannot rule out 
the possibility that we have been extremely unlucky and 
are just observing  
a large statistical fluctuation. 

It is important to find out whether the Westphal fields
are outlyers or, conversely, if the observed dispersion 
in the number-counts reflects the real dispersion
between WFPC2 fields. If the latter is true, 
the size of the errors severely hampers 
the precision of extinction measurements derived from
galaxy number-counts.  Also, studies of the clustering
properties of a number of deep WFPC2 fields should determine whether
the inadequacy of the estimated clustering errors is real
or caused by an overly optimistic completeness limit.  
Finally, even if it sounds improbable, 
there is of course the possibility  
that some obscuring material lies in between us and the
Westphal fields.  It would have to be colder than $\sim$ 10-15 K
to not show up in the 100 $\mu$m maps (Schlegel et al.\ 1998), 
but this is not unlikely
at high Galactic latitude, far from the interstellar UV radiation field; 
it would also have to leave the colors of the galaxies unchanged.
The answer to the number-count discrepancy is beyond the scope of this paper,
but it will be very interesting, either way.

\acknowledgements
We acknowledge Max Mutchler for assistance with the reduction of the 
GRB WFPC2 images.
We are indebted to Ron Allen and Harry Ferguson for their
very useful comments.
We thank Dave Schlegel for providing measurements of the
reddening towards all the fields we use in this paper,
and Paul Goudfrooij for help with {\em synphot}.
Support for this work was provided partly by NASA
through grant AR-06400.01-95A and partly by the Director's
Discretionary Research Fund at the Space Telescope Science Institute.

\newpage
\twocolumn
 
\vfill
\eject

\newpage
\onecolumn
\tighten
\vfill

\clearpage
\eject

\begin{table}[tp]
\vspace{-4in}
\hspace*{-0.8in}\psfig{figure=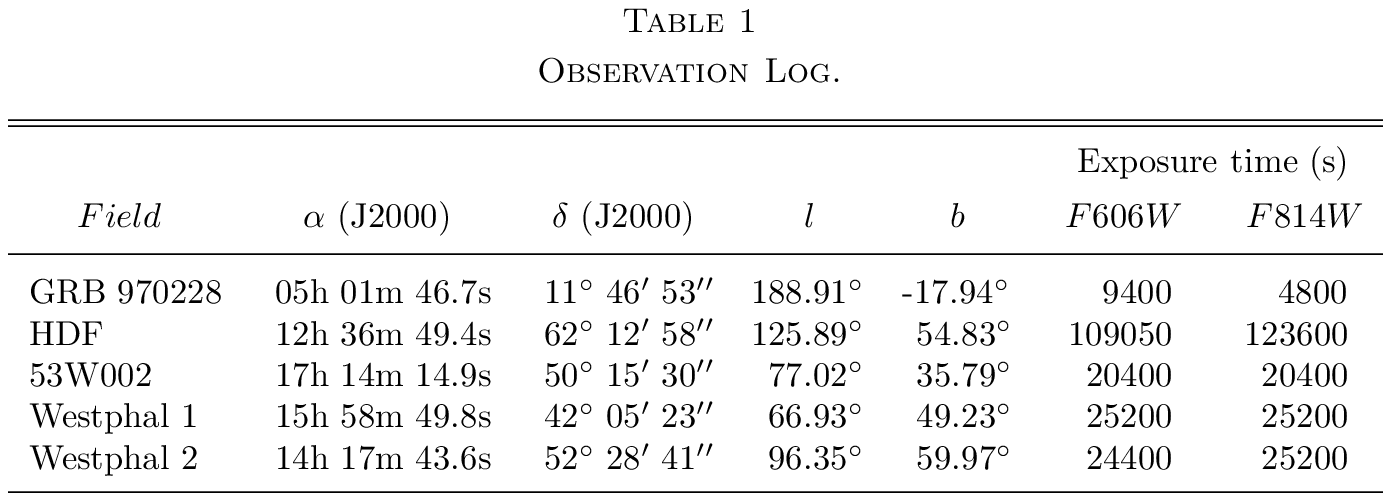,width=8in,angle=0}
\label{tobs}
\end{table}

\clearpage
\eject

\begin{table}[tp]
\vspace{-0.9in}
\hspace*{-0.3in}\psfig{figure=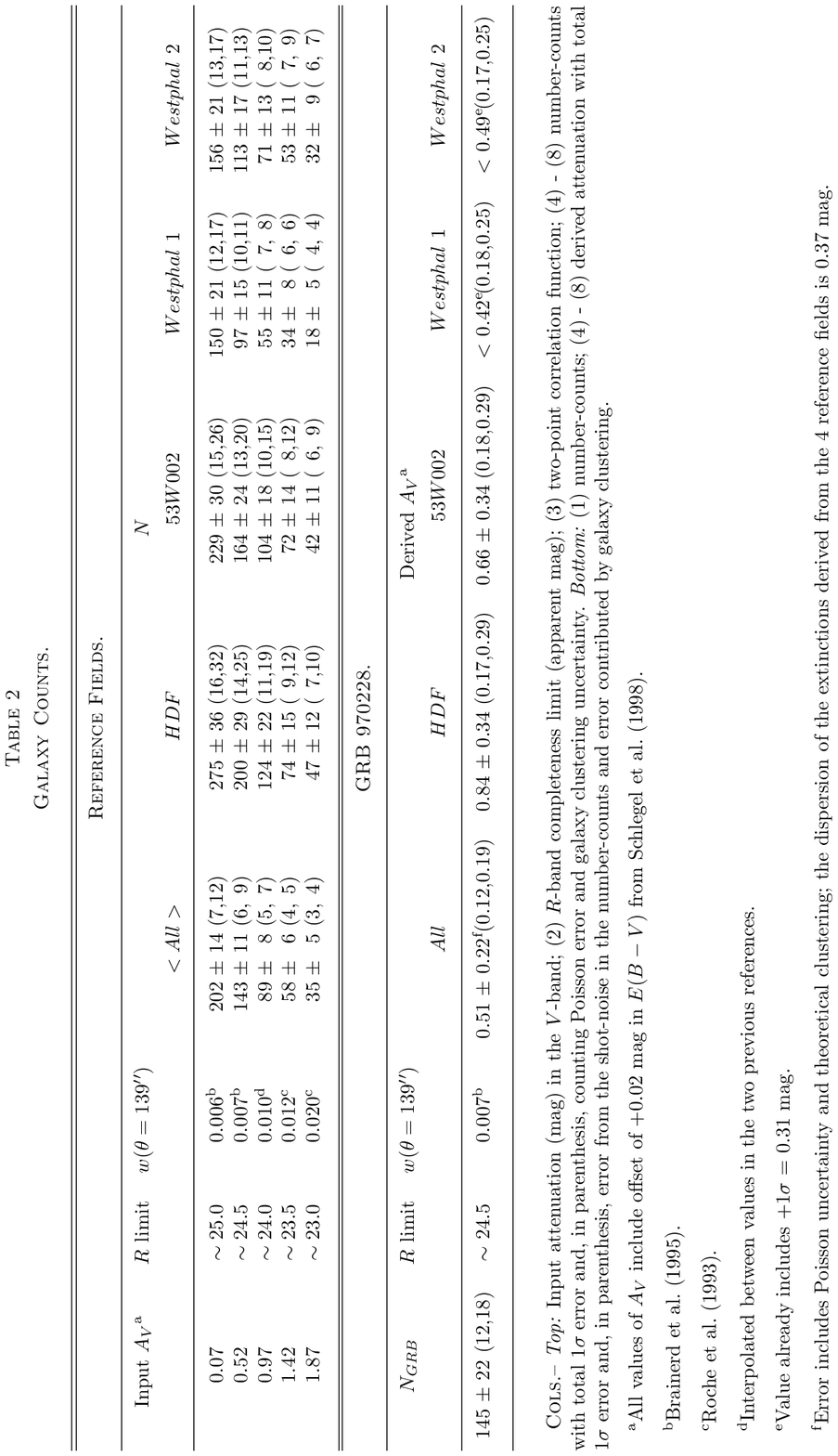,width=7.5in,angle=0}
\label{tcounts}
\end{table}

\clearpage
\eject

\begin{table}[tp]
\vspace{-0.9in}
\hspace*{-0.3in}\psfig{figure=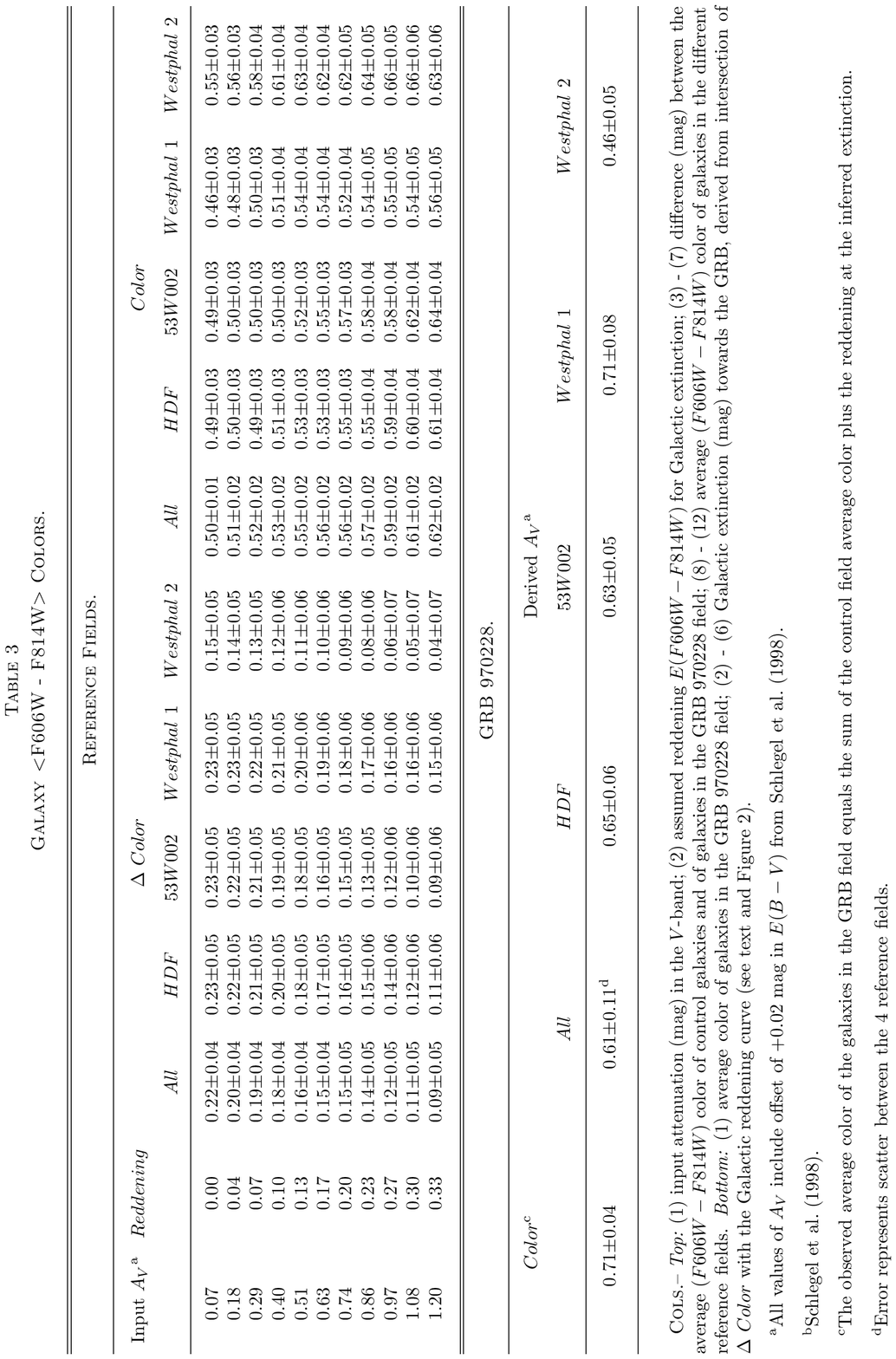,width=7.5in,angle=0}
\label{tcolors}
\end{table}

\clearpage
\eject

\clearpage
\eject

\end{document}